# Tuning the carrier concentration to improve the thermoelectric performance of CuInTe$_2$ compound

J. Wei, H. J. Liu[*], L. Cheng, J. Zhang, J. H. Liang, P. H. Jiang, D. D. Fan, J. Shi

*Key Laboratory of Artificial Micro- and Nano-structures of Ministry of Education and School of Physics and Technology, Wuhan University, Wuhan 430072, China*

The electronic and transport properties of CuInTe$_2$ chalcopyrite are investigated using density functional calculations combined with Boltzmann theory. The band gap predicted from hybrid functional is 0.92 eV, which agrees well with experimental data and leads to relatively larger Seebeck coefficient compared with those of narrow-gap thermoelectric materials. By fine tuning the carrier concentration, the electrical conductivity and power factor of the system can be significantly optimized. Together with the inherent low thermal conductivity, the *ZT* values of CuInTe$_2$ compound can be enhanced to as high as 1.72 at 850 K, which is obviously larger than those measured experimentally and suggests there is still room to improve the thermoelectric performance of this chalcopyrite compound.

Thermoelectric (TE) materials have attracted great interest recently due to their great potential in directly converting heat into electricity or vice versa. The performance of a thermoelectric material at temperature $T$ can be characterized by the dimensionless figure of merit:

$$ZT = S^2 \sigma T / (\kappa_e + \kappa_L), \qquad (1)$$

which includes the Seebeck coefficient $S$, the electrical conductivity $\sigma$, the electronic thermal conductivity $\kappa_e$, and the lattice thermal conductivity $\kappa_L$. In principle, the *ZT* value can be improved by utilizing some strategies to increase the power factor ($S^2\sigma$) and/or decrease the thermal conductivity ($\kappa_e + \kappa_L$). However, it is usually very challenging to do so because of strong interdependence of these transport coefficients. Over the past several decades, major effort has been directed towards the optimization of current TE materials such as $Bi_2Te_3$, SiGe alloy, PbTe and their doped compounds, as well as the exploration of novel high-performance compounds [1, 2, 3, 4, 5, 6, 7, 8]. Usually, the state-of-the-art TE materials with higher *ZT* values are limited to only a few narrow-gap semiconductors and most of them were developed in the 1960s [9]. In order to further improve the TE performance, we need to explore new materials such as chalcopyrites $CuGaTe_2$ [10, 11] and $AgGaTe_2$ [12, 13]. This kind of compounds were found to exhibit larger *ZT* value at high temperature region, which can be explained by significant decrease in the thermal conductivity at increased temperature.

---

[*] Author to whom correspondence should be addressed. Electronic mail: phlhj@whu.edu.cn

As a typical chalcopyrite compound, the thermoelectric performance of $CuInTe_2$ has attracted much attention recently since it inherently has larger Seebeck coefficients and lower thermal conductivity. Liu *et al.* [14] reported that $CuInTe_2$ is a promising thermoelectric material with a larger *ZT* value of 1.18 at 850 K. Using chemical vapor transport (CVT) technique, Prabukanthan *et al.* synthesized single crystal $CuInTe_2$ [15]. The measured band gap is 1.04 eV, which is much larger than those of traditional thermoelectric materials (for example, 0.11 eV for $Bi_2Te_3$ [16] and 0.14 eV for $Sb_2Te_3$ [17]). It should be mentioned that larger band gap usually leads to relatively small electrical conductivity compared with those of state-of-the-art materials. In this regard, tuning carrier concentration through element doping could be an effective way to optimize both the electrical conductivity ($\sigma$) and the power factor ($S^2\sigma$). Cheng *et al.* [18] reported that the carrier concentration of $CuInTe_2$ is greatly enhanced by doping the system with Cd, which leads to an enhancement of *ZT* value by more than 100% at room temperature and around 20% at 600 K. In addition, Kosuga *et al.* [19] found that the reduction in the Cu content acts as hole doping and the sample of $Cu_{1-x}InTe_2$ with x =0.1 exhibits the largest *ZT* value of 0.54 at 710 K. It should be mentioned that all these are experimental works and the reported *ZT* values are very different from each other, ranging from 0.40 to 1.18 at high temperature region. Moreover, the samples obtained in different experiments exhibit quite different carrier concentration, which can have significant influence on the electronic transport properties of $CuInTe_2$. It is thus very necessary to conduct an extensive theoretical investigation on the electronic and transport properties of $CuInTe_2$, which may shed some light on further optimization of the thermoelectric performance of this chalcopyrite compound.

Our theoretical approach combines the first-principles calculations and Boltzmann transport theory. The structure optimization and electronic properties of $CuInTe_2$ are calculated by using projector augmented-wave (PAW) method [20] as implemented in the Vienna *ab initio* simulation package (VASP) [21, 22, 23]. We use the hybrid functional in the form of Heyd-Scuseria-Ernzerhof (HSE). The Brillouin zone is sampled by a $7 \times 7 \times 4$ **k**-mesh and a plane wave cutoff energy of 350 eV is adopted in the calculations. The system is fully relaxed until the magnitude of the forces acting on all the atoms become less than 0.01 eV/Å. The electronic transport coefficients are derived by using the semi-classical Boltzmann theory [24], where the carrier concentration and temperature dependence of relaxation time is obtained by fitting the existing experimental data.

The crystal structure of $CuInTe_2$ is shown in Figure 1, which exhibits typical chalcopyrite structure (space group I-42d) with 16 atoms per unit cell. Each Cu and In atom is connected by 4 Te atoms, forming diamond-like structure. The experimentally measured lattice constants of $CuInTe_2$ are $a = b = 6.20$ Å and $c = 12.44$ Å [15], which are exclusively used in the following calculations. It is found that the Cu and In atoms form ordered long-range cubic framework while the Te atoms form localized short-range non-cubic lattice distortions. Such special structure characteristic can block the heat transport while has less effect on the electron conduction, which may be very beneficial to the thermoelectric performance of $CuInTe_2$, as well as other chalcopyrite compounds [25].

Since the standard density functional theory (DFT) with the local-density approximation (LDA) or generalized gradient approximation (GGA) underestimates the band gap of chalcopyrite compounds [11, 26], we use the HSE hybrid functional to calculate the electronic properties of $CuInTe_2$ compound. Figure 2(a) displays the energy band structure of the compound along several high symmetry lines in the irreducible Brillouin zone. Our calculations indicate that the band gap of CuInTe$_2$ is 0.92 eV, which is consistent with the experimental results of 0.96~1.06 eV [15, 27, 28, 29]. The conduction band minimum (CBM) and the valence band maximum (VBM) are both located at the Γ point, which is a common characteristic to the chalcopyrite materials. It should be noted that there are highly degenerate energy bands near the VBM (see circled area of Figure 2(a)), which can simultaneously leads to large Seebeck coefficient and high electrical conductivity [1]. This observation indicates that the *p*-type CuInTe$_2$ could perhaps have better thermoelectric performance than *n*-type system. In order to further understand the electronic properties of CuInTe$_2$, we calculate the density of states (DOS) of the compound as shown in Figure 2(b). It can be seen that the conduction band edge is composed primarily of the states from In and Te atoms, while the contribution from the Cu atom is smaller. However, for the valence band edge, it mostly consists of Cu and Te states and they are effectively hybridized with each other. Compared with the case of conduction band edge, In atoms no longer play an important role in increasing the DOS at VBM. If In atoms can be substituted by atoms with less valence electrons, one can increase the hole concentration and thus the electrical conductivity while have less effect on the Seebeck coefficient, and overall leads to improved power factor. Indeed, Cheng *et al.* [18] reported that due to the substitution of Cd at In sites, the carrier concentration is greatly increased, leading to greatly enhanced electrical conductivity and power factor.

Based on the band structure, we are able to evaluate the transport coefficients by using the semi-classical Boltzmann theory [24] and the rigid-band approach [30]. Figure 3(a) shows the calculated Seebeck coefficients $S$ of CuInTe$_2$ as a function of chemical potential $\mu$ at 300 K. Note here the chemical potential indicates the doping level or carrier concentration of the system, and the positive and negative $\mu$ indicate *n*-type and *p*-type carriers, respectively. We see that the Seebeck coefficients of CuInTe$_2$ exhibit two obvious peaks around the Fermi level $(\mu=0)$. The absolute values of them (~1600 $\mu$V/K) are much larger than those of most conventional thermoelectric materials due to relatively larger band gap of CuInTe$_2$. Unlike the Seebeck coefficient which is independent of the relaxation time $\tau$, the electrical conductivity $\sigma$ can only be calculated with respect to $\tau$. The room temperature $\sigma/\tau$ as a function of chemical potential $\mu$ is plotted in Figure 3(b). We see there is a sharp increase of $\sigma/\tau$ around the band edge and it is more pronounced for the *p*-type system, which may be benefited from the band degeneracy at the VBM. In order to evaluate the particular value of $\sigma$, we need to know the relaxation time which is usually very complicated to calculate since it depends on the detailed scattering mechanism involved. For simplicity, here the relaxation time is obtained by fitting the experimentally measured electrical conductivity [19, 31], and its carrier concentration and temperature dependence can be expressed as:

$$\tau = AT^{-1}n^{-1/3}, \qquad (2)$$

Note such treatment of relaxation time has been generally used for other thermoelectric materials such as CuGaTe$_2$, AgGaTe$_2$ and ZnO system [11, 13, 32]. For the CuInTe$_2$ compound, the averaged value of constant *A* in Equation (2) is calculated to be 2.47×10$^{-5}$ sK/cm. Figure 3(c) plots the relaxation time as a function of chemical potential, which exhibits large value around the Fermi level for both electrons and holes and can reach as high as 10$^{-12}$ s. Inserting the relaxation time into Figure 3(b), one can expect quite larger electrical conductivity in the order of magnitude of 10$^7$ S/m. Together with much larger Seebeck coefficient shown in Figure 3(a), this finding suggests very favorable thermoelectric performance of CuInTe$_2$ at particular chemical potential or carrier concentration. For the electronic thermal conductivity $\kappa_e$, we use the Wiedemann-Franz law:

$$\kappa_e = L\sigma T, \qquad (3)$$

where the Lorenz number $L$ can be expressed as:

$$L(\eta) = \left(\frac{\kappa_B}{e}\right)^2 \left( \frac{(r+7/2)F_{r+5/2}(\eta)}{(r+3/2)F_{r+1/2}(\eta)} - \left( \frac{(r+5/2)F_{r+3/2}(\eta)}{(r+3/2)F_{r+1/2}(\eta)} \right)^2 \right), \qquad (4)$$

Here $\eta$ is the reduced Fermi energy, and $r$ is the scattering parameter which is $-1/2$ for acoustic-phonon scattering. The Fermi integral $F_i(\eta)$ in Equation (4) is given by:

$$F_i(\eta) = \int_0^\infty \frac{x^i dx}{e^{x-\eta}+1}, \tag{5}$$

The calculated Lorenz number of $CuInTe_2$ is $1.52 \sim 1.71 \times 10^{-8}$ $V^2/K^2$ in the carrier concentration range from $1.0 \times 10^{19}$ to $1.0 \times 10^{20}$ $cm^{-3}$, which is basically a constant and indicates that the electronic thermal conductivity should exhibit similar behavior as that of electrical conductivity and thus is not shown here. Inserting all these transport coefficients into Equation (1) and using a room temperature lattice thermal conductivity of 4.50 W/mK fitted from experimental data [33], we are now able to evaluate the thermoelectric performance of $CuInTe_2$ compound. Figure 3(d) plots the room temperature *ZT* value as a function of chemical potential. We see there are two remarkable peaks around the Fermi level, which is 0.15 at optimized concentration of $2.69 \times 10^{19}$ $cm^{-3}$ for the *p*-type $CuInTe_2$, and 0.10 for the *n*-type system at carrier concentration of $9.96 \times 10^{17}$ $cm^{-3}$. The relatively larger *ZT* value of *p*-type $CuInTe_2$ is due to the fact that *p*-type system has larger electrical conductivity compared with *n*-type system while the Seebeck coefficients and electronic thermal conductivities of them are similar at optimized carrier concentration.

We now discuss the temperature dependence of the thermoelectric performance. Figure 4 plots the lattice thermal conductivity of $CuInTe_2$ as a function of temperature (solid line), which is obtained by fitting the available experimental data [33] at different temperatures (black squares) and can be expressed by:

$$\kappa_L = 1729.14/T - 1.27. \tag{6}$$

It can be seen that the $\kappa_L$ of $CuInTe_2$ decreases quickly from 4.0 to 0.94 W/mK as the temperature increases from 300 to 700 K. The extremely low thermal conductivity at higher temperature shows its great potential as TE material. In Figure 4, we also plot the calculated *ZT* value of *n*-type and *p*-type $CuInTe_2$ as a function of temperature at optimal carrier concentration. For the *p*-type system, the *ZT* value can be significantly enhanced from 0.15 to 1.72 as the temperature increases from 300 to 850 K. The optimum carrier concentration is in the range from $2.69\times10^{19}$ cm$^{-3}$ to $4.25\times10^{19}$ cm$^{-3}$, which is larger than that found in the experiment [14]. It is thus reasonable to expect that if the carrier concentration can be fine tuned, the *ZT* value of CuInTe$_2$ compound can be further improved. On the other hand, the temperature dependence of *n*-type CuInTe$_2$ is not as strong as *p*-type system, and the maximum *ZT* value is only 0.87 at 850 K where the carrier concentration is relatively smaller ($1.48\times10^{18}$ cm$^{-3}$). Our calculated *ZT* values are summarized in Table I where the corresponding optimal carrier concentration and transport coefficients are also given. It should be mentioned that the *ZT* value of 1.72 for *p*-type CuInTe$_2$ predicted in the present work not only significantly exceeds the experimental results, but is also higher than those of other *p*-type chalcopyrite compounds reported so far. Our theoretical predictions thus suggest that CuInTe$_2$ is a very promising TE material operated at high temperature region, and there is still room to improve the thermoelectric performance of this chalcopyrite compound by further optimizing the carrier concentration.

In summary, our theoretical calculations demonstrate that CuInTe$_2$ could be optimized to exhibit very good thermoelectric performance by fine tuning the carrier concentration. The predicted *p*-type *ZT* value is larger than that of *n*-type, and a maximum *ZT* value of 1.72 can be achieved at 850 K with optimum concentration of $4.25\times10^{19}$ cm$^{-3}$. Such a concentration can be in principle realized by substituting In atoms with atoms such as Cd, Ag and Pd. Taking Cd doping as an example, we find that CuIn$_{1-x}$Cd$_x$Te$_2$ compounds with x=0.02 can fulfill such requirement. Our calculations provide a detailed understanding of the excellent thermoelectric performance of CuInTe$_2$ and may shed some light on further enhancing the thermoelectric performance of this chalcopyrite compound.

We thank financial support from the National Natural Science Foundation (Grant No. 51172167) and the "973 Program" of China (Grant No. 2013CB632502).

**Table I** Optimized *p*- and *n*-type *ZT* values of CuInTe$_2$ compound at different temperature. The corresponding carrier concentration and transport coefficients are also indicated.

| T (K) | μ (eV) | p/n (cm$^{-3}$) | S (μV/K) | σ (10$^4$ S/m) | S$^2$σ (mW/mK$^2$) | κ$_e$ (W/mK) | κ$_L$ (W/mK) | ZT |
|---|---|---|---|---|---|---|---|---|
| 300 | −0.44 | 2.69×10$^{19}$ | 245 | 3.85 | 2.32 | 0.18 | 4.50 | 0.15 |
| 400 | −0.44 | 3.42×10$^{19}$ | 261 | 3.42 | 2.32 | 0.21 | 3.06 | 0.28 |
| 500 | −0.43 | 3.95×10$^{19}$ | 275 | 3.03 | 2.29 | 0.23 | 2.19 | 0.47 |
| 600 | −0.41 | 4.33×10$^{19}$ | 289 | 2.68 | 2.24 | 0.25 | 1.62 | 0.72 |
| 700 | −0.39 | 4.44×10$^{19}$ | 306 | 2.32 | 2.17 | 0.25 | 1.20 | 1.05 |
| 800 | −0.36 | 4.37×10$^{19}$ | 324 | 1.98 | 2.08 | 0.24 | 0.90 | 1.46 |
| 850 | −0.35 | 4.25×10$^{19}$ | 334 | 1.82 | 2.03 | 0.23 | 0.77 | 1.72 |
| 300 | 0.51 | 9.96×10$^{17}$ | −247 | 2.60 | 1.59 | 0.12 | 4.50 | 0.10 |
| 400 | 0.50 | 9.65×10$^{17}$ | −260 | 1.93 | 1.31 | 0.12 | 3.06 | 0.16 |
| 500 | 0.48 | 9.71×10$^{17}$ | −271 | 1.53 | 1.13 | 0.12 | 2.19 | 0.24 |
| 600 | 0.46 | 1.04×10$^{18}$ | −280 | 1.30 | 1.02 | 0.12 | 1.62 | 0.35 |
| 700 | 0.45 | 1.18×10$^{18}$ | −288 | 1.15 | 0.96 | 0.12 | 1.20 | 0.50 |
| 800 | 0.43 | 1.37×10$^{18}$ | −297 | 1.05 | 0.92 | 0.13 | 0.90 | 0.72 |
| 850 | 0.43 | 1.48×10$^{18}$ | −303 | 1.00 | 0.92 | 0.13 | 0.77 | 0.87 |

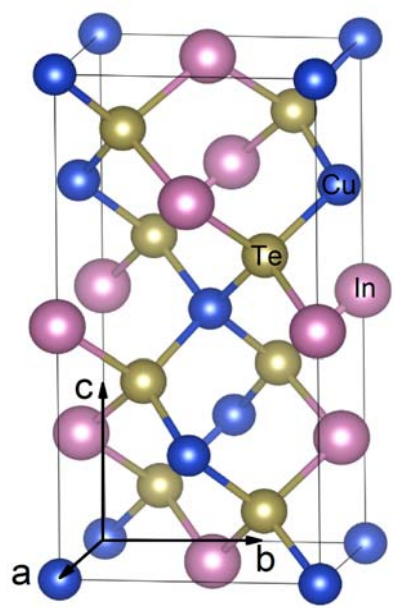

**Figure 1** The ball-and-stick model of CuInTe$_2$ compound.

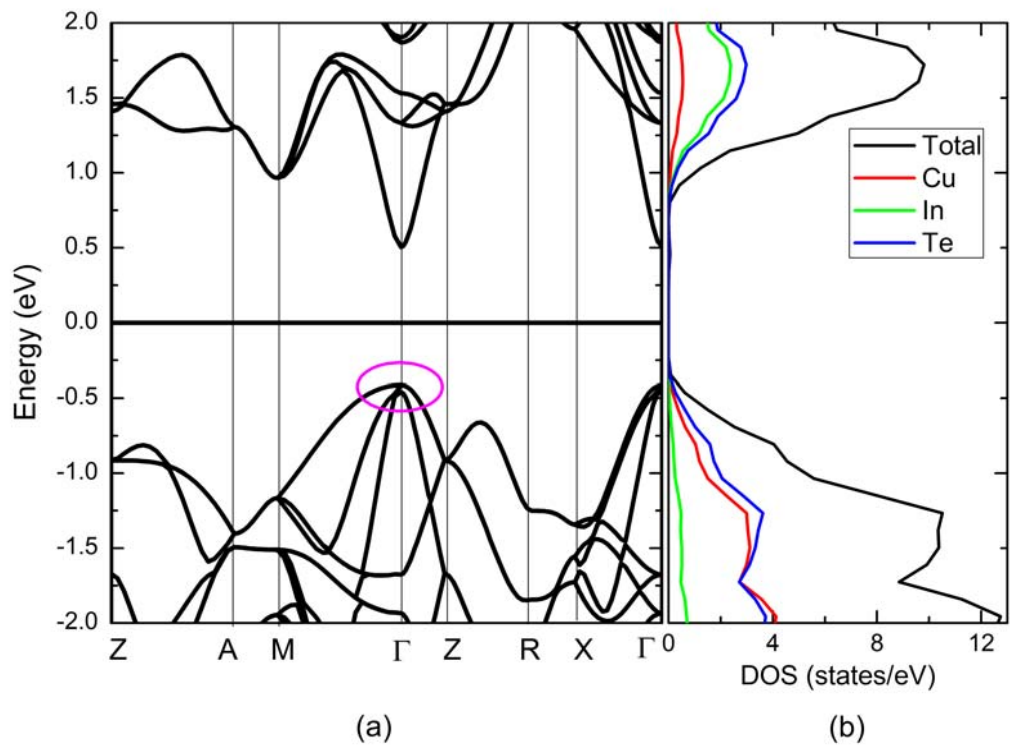

**Figure 2** Energy band structures and DOS of the CuInTe$_2$ compound. The Fermi level is at 0 eV.

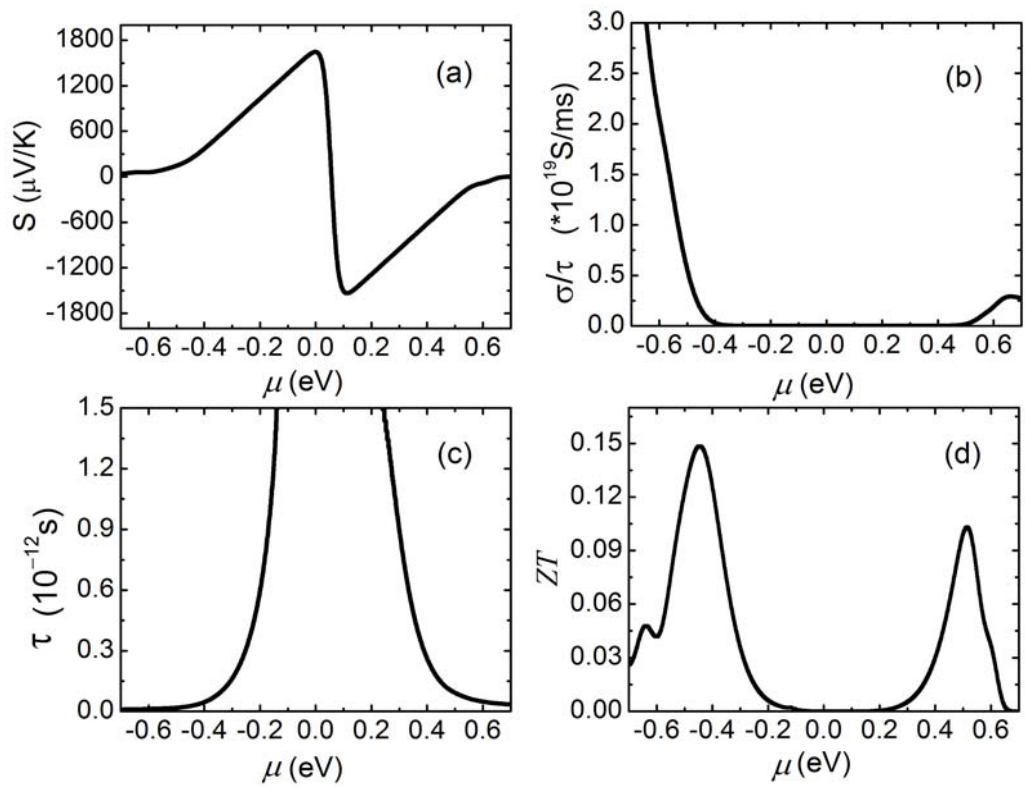

**Figure 3** Calculated room temperature transport coefficients as a function of chemical potential for the $CuInTe_2$ compound: (a) the Seebeck coefficient, (b) the electrical conductivity with respect to the relaxation time, (c) the relaxation time, and (d) the *ZT* value.

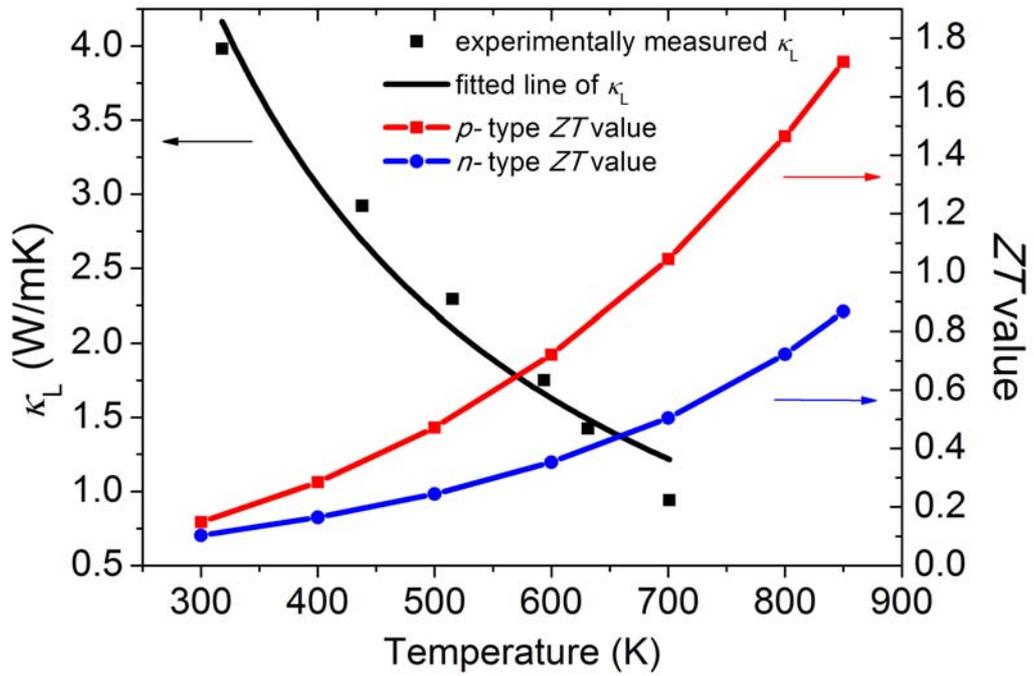

**Figure 4** The temperature dependence of lattice thermal conductivity and *ZT* value for the CuInTe$_2$ compound.